\documentstyle[12pt,graphicx,here]{article}

\begin{document}

\begin{titlepage}
\def\thefootnote{\fnsymbol{footnote}}       

\begin{center}
\mbox{ } 

\vspace*{-3cm}

{\Large \mbox{\hspace{-1.3cm} EUROPEAN ORGANIZATION FOR NUCLEAR RESEARCH}} 

\end{center}
\vskip 0.5cm
\begin{flushright}
\large
hep-ex/9710022
\end{flushright}
\begin{center}
\vskip 0.5cm
{\Huge\bf
Charged Higgs Boson \\
Detector Aspects}
\vskip 1.0cm
{\Large\bf Andr\'e Sopczak}\\
\smallskip
\large DESY -- Zeuthen
\footnote{\normalsize now: University of Karlsruhe  ~~~~
e-mail: andre.sopczak@cern.ch}

\vskip 1.5cm
\centerline{\large \bf Abstract}
\end{center}

\renewcommand{\baselinestretch} {1.2}

\large
The discovery potential for charged Higgs bosons has been studied
including background simulations for $\sqrt{s} = 500$~GeV and 
a luminosity of 10~fb$^{-1}$
depending on the hadronic calorimeter resolution of a LC500 detector.
The hadronic decay channel $H^+ H^- \rightarrow c\bar{s}\bar{c}s$ 
is in particular suited for this study due to a large sensitivity 
dependence on the hadron calorimeter performance.
For this study a hadronic energy resolution as achieved at LEP 
experiments is sufficient to reach a very good sensitivity. 
A calorimeter with weaker energy resolution 
would reduce the sensitivity by a  factor of two.
\renewcommand{\baselinestretch} {1.}

\normalsize
\vspace{1cm}
\begin{center}
{\sl
Contribution to the proceedings of the ``ECFA/DESY Study on Physics and 
Detectors for the Linear Collider'', DESY 97-123E, ed. R.~Settles
}
\end{center}
\vfill
\end{titlepage}


\begin{center}
{\Large \bf Charged Higgs Boson Detector Aspects} \\
\bigskip
{\large Andr\'e Sopczak} \\
\bigskip
{\it DESY -- Zeuthen}\footnote{\normalsize 
now: University of Karlsruhe~~~~~~e-mail: andre.sopczak@cern.ch   }
\end{center}

\begin{abstract}
\normalsize 
The discovery potential for charged Higgs bosons has been studied
including background simulations for $\sqrt{s} = 500$~GeV and 
a luminosity of 10~fb$^{-1}$
depending on the hadronic calorimeter resolution of a LC500 detector.
The hadronic decay channel $H^+ H^- \rightarrow c\bar{s}\bar{c}s$ 
is in particular suited for this study due to a large sensitivity 
dependence on the hadron calorimeter performance.
For this study a hadronic energy resolution as achieved at LEP 
experiments is sufficient to reach a very good sensitivity. 
A calorimeter with weaker energy resolution 
would reduce the sensitivity by a  factor of two.
\end{abstract}

A discovery of charged Higgs bosons would be an unambiguous evidence
that the electro\-weak symmetry-breaking sector of the Standard Model
consists of at least two Higgs doublets.
Charged Higgs bosons are pair produced at a LC500 if kinematically allowed. 
In the hadronic decay channel $H^+H^- \rightarrow c\bar{s}\bar{c}s$, 
the selection procedure depends largely on the mass resolution.
In particular, this decay channel is rather independent of 
the b-tagging capabilities of the detector which are very 
important for neutral Higgs boson sensitivity. 
Thus, this charged Higgs boson hadron channel is 
suited to study effects of the hadronic calorimeter 
detector resolution.

This analysis investigates the influence of the detector 
performance of a hadronic calorimeter on the expected detection
efficiency for $H^+H^- \rightarrow c\bar{s}\bar{c}s$. 
The analysis is based on previous charged Higgs boson 
studies~[1] for a 500~GeV linear collider.
The detector effects of hadronic events can be studied efficiently
with a pa\-ra\-me\-tric detector simulation which takes into account
the detector geometry and the intrinsic resolutions.
Most important for the Higgs search in the multi-jet channels are the
energy and spatial resolutions in the central part of the detector
($| \cos\theta | < 0.7$)
because of the approximate $\sin^2\theta$ distribution of the 
production angle for the signal.
Typical resolutions are approximated by smearing single particle
energies and directions as listed in
Table~\ref{tab:lthreeresolutions}.
The resulting resolutions are in agreement with measurements
at the Z resonance and give confidence in the simulations at higher 
energies.

\begin{table}[htb]
\vspace*{-0.4cm}
\renewcommand{\arraystretch}{1.2}
\begin{center}
\begin{tabular}{|c|c|c|}\hline
Particle  & Energy Resolution (in \%) & Spatial Resolution \\ \hline\hline
e,$\gamma$
          & $\Delta E / E \approx 5 / \sqrt{E} $
          & $\Delta\phi,~\Delta\theta\approx 10$~mrad  \\ \hline
Single Hadrons  
          & $\Delta E / E \approx 55 / \sqrt{E} + 5 $
          & $\Delta\phi,~\Delta\theta\approx {150}~{\mathrm{mrad}} / \sqrt{E}$ \\ \hline
\end{tabular}
\vspace*{-0.2cm}
\caption{  Smearing used for the simulation of detector effects
           in the LC500 Higgs search. 
           $E$ is the energy in GeV of electrons, photons,
           or single hadrons.}
\label{tab:lthreeresolutions}
\end{center}
\vspace*{-0.4cm}
\end{table}

A previous $H^+H^-$ analysis~[1] had been performed
using a detector simulation corresponding to the L3 detector specifications.
With an event-shape preselection, the background
can be reduced to about 1900 $t\bar{t}$, 1600 $q\bar{q}$, 140 $W^+W^-$, 
and 100 $ZZ$ events, with about 50\% signal selection efficiency.
Most $q\bar{q}$ background is reduced by comparison of
reconstructed invariant masses, since for the charged Higgs bosons 
both pairs are expected to have identical invariant masses.
$ZZ$ and $W^+W^-$ background is reduced by rejection of events where
$Z$ and $W$ masses were reconstructed. Independent of the hadronic
resolution, top-quark background is further reduced by anti-b-tagging.
Detailed selection cuts are given in~[1].
This analysis keeps the selection cuts unchanged.  
The same detector simulation is applied, except that a 
hadronic resolution factor $f$ is introduced:
$$
\Delta E / E = f \times (55/\sqrt{E} + 5)\%. 
$$

\begin{figure}[H]
\vspace*{0.5cm}
\begin{center}
\includegraphics[width=12cm,clip]{%
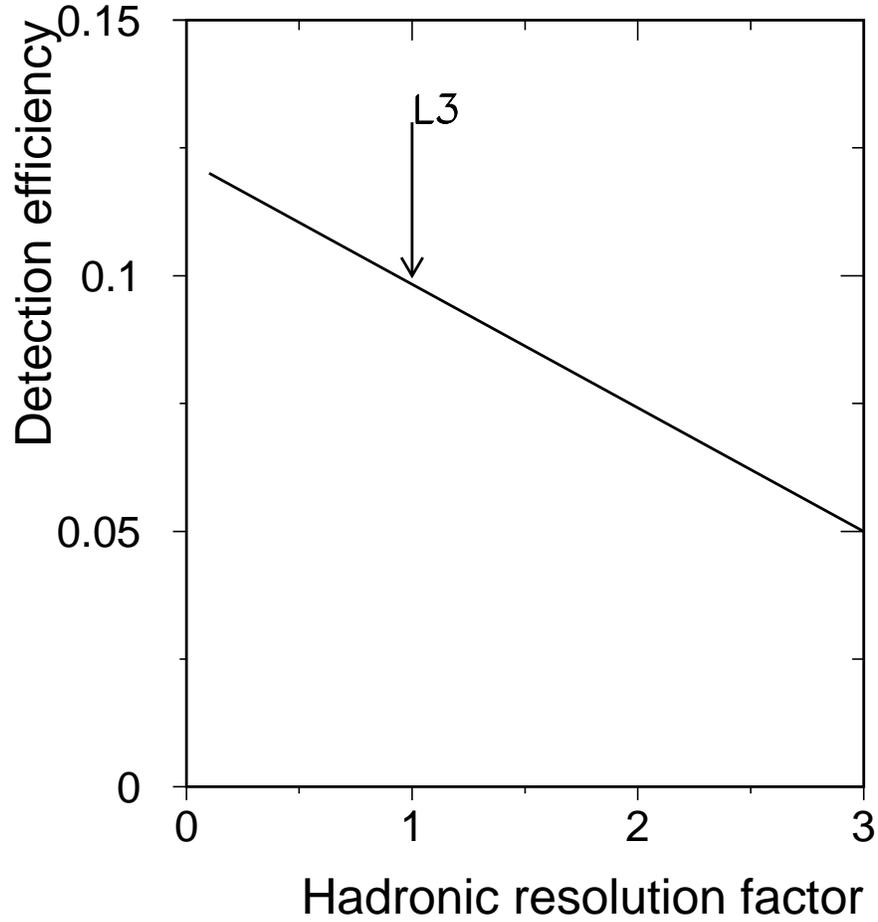}
\end{center}

\caption
{\label{fig:had}
Detection efficiency for the process $H^+H^- \rightarrow c\bar{s}\bar{c}s$
with a 180~GeV charged Higgs mass as a function of a 
scale factor for the hadron calorimeter resolution. 
The expected detection efficiency assuming the L3 detector resolution 
is indicated.}
\end{figure}

The variation of the $H^+H^- \rightarrow c\bar{s}\bar{c}s$ 
detection efficiency as a 
function of the hadronic resolution factor is shown in 
Fig.~\ref{fig:had}. 
The detection efficiency
in the hadronic $H^+H^-$ channel depends significantly on 
the hadronic calorimeter resolution. Sensitivity can more than
double compared to a less precise hadron calorimeter. For example,
if the hadronic resolution improves from $f=3$ to $f=1$, the 
detection efficiency improves from about 5 to 10\%.
On the other hand, an unrealistic detector resolution ten 
times as good as a LEP experiment would 
only increase the detection efficiency from about 10 to 12\%
for a 180~GeV charged Higgs boson.
Concluding, a hadronic calorimeter resolution already achieved
at the LEP experiments suffice for a performing sensitivity.

\clearpage
\noindent
{\bf References}

[1] A.~Sopczak,
DESY 93-123C p. 121; Z.~Phys. {\bf C65} (1995) 449.

\end{document}